\documentclass[twocolumn,preprintnumbers,amsmath,amssymb]{revtex4}
\usepackage{epsfig}
\usepackage{dcolumn}
\usepackage{bm}

\newcommand{\be}{\begin{equation}}
\newcommand{\ee}{\end{equation}}
\newcommand{\bea}{\begin{eqnarray}}
\newcommand{\eea}{\end{eqnarray}}

\newcommand{\btab}{\begin{tabbing}}
\newcommand{\etab}{\end{tabbing}}

\newcommand{\bit}{\begin{itemize}}
\newcommand{\eit}{\end{itemize}}

\newcommand{\ben}{\begin{enumerate}}
\newcommand{\een}{\end{enumerate}}

\newcommand{\bfig}{\begin{figure}[h]}
\newcommand{\efig}{\end{figure}}

\newcommand{\hs}[1]{\hspace*{#1}}

\newcommand{\LB}[1]{\label{#1}}

\begin{document}

\title{Geometric Diagnostics of Complex Patterns: Spiral Defect Chaos}
\author{Hermann Riecke and Santiago Madruga\footnote{present
address: Max-Planck-Institute for Physics of Complex Systems, D-01187
Dresden, Germany}}
\affiliation{Engineering Science and Applied Mathematics, 
Northwestern University, Evanston, IL 60208, USA}
\date{\today}

\begin{abstract}
 
Motivated by the observation of spiral patterns in a wide range of
physical, chemical, and biological systems we present an approach that
aims at characterizing quantitatively spiral-like elements in complex
stripe-like patterns. The approach provides the location of the spiral
tip and the size of the spiral arms in terms of their arclength and
their winding number. In addition, it yields the number of pattern
components (Betti number), as well as their size and certain aspects
of their shape. We apply the method to spiral defect chaos in
thermally driven Rayleigh-B\'enard convection and find that the
arclength of spirals decreases monotonically with decreasing Prandtl
number of the fluid and increasing heating. By contrast, the winding
number of the spirals is non-monotonic in the heating. The
distribution function for the number of spirals is significantly
narrower than a Poisson distribution. The distribution function for
the winding number shows approximately an exponential decay. It
depends only weakly on the heating, but strongly on the Prandtl
number. Large spirals arise only for larger Prandtl numbers ($Pr
\gtrsim 1$). In this regime the joint distribution for the spiral
length and the winding number exhibits a three-peak structure
indicating the dominance of Archimedean spirals of opposite sign and
relatively straight sections. For small Prandtl numbers the
distribution function reveals a large number of small compact pattern
components.

\end{abstract} 

\maketitle

{\bf  Many systems in nature exhibit complex pattern that may be
stationary or time-dependent, possibly in a chaotic fashion. To
understand these patterns and any transition they might undergo it is
important to have {\it quantitative} measures that characterize the
relevant properties of the patterns and their time dependence. Due to
the multitude of different types of patterns it is not to be expected
that a single measure would be sufficient to capture the qualitatively
different aspects of the various patterns. Thus, while quite a few
different measures have been developed over the years, so far no
convincing approach is available that extracts the characteristic
features of patterns dominated by spiral-like pattern components. This
paper presents a method that allows to determine quantitatively
features like the length of a spiral and its winding number. The
approach is, however, not limited to proper spirals and yields
additional insightful measures.}

\section{Introduction}

Complex spatio-temporal patterns abound in nature. Many of them
contain also chiral elements like spirals or spiral segments.
Particularly well-known and beautiful examples are spiral waves in
oscillating and excitable chemical systems
\cite{BaNe95,OuFl96,BeOu97,GuLi03,LiHa04}, chiral structures in microbial
growth patterns \cite{Sh95,LeBe04}, and spiral defect chaos in thermally
driven convection \cite{MoBo93,BoPe00}. Other examples are spirals
arising in vibrationally excited waves on the surface of granular
material \cite{BrLe01}, in calcium waves in oocytes \cite{LeGi91a} and
hippocampus \cite{HaZa98}, and in the aggregation patterns of the
amoeba Dictyostelium \cite{PaLe97,DoKi01}. These patterns often evolve
chaotically in time and arise in transitions from simpler structures
and may in turn undergo further transitions to other, possibly more
complex structures. To gain detailed understanding of such
spatio-temporally chaotic states and their transitions it is important
to have {\it quantitative} measures that extract the relevant features
of the patterns. 

Reflecting the variety of spatio-temporal structures observed, a host
of different measures and approaches have been developed over the
years to characterize the various types of spatio-temporal chaos.
Global measures for the disorder in the pattern can be obtained from
the correlation function and the spectral entropy \cite{CaEg97}. These
measures are quite general; they do not give any local information and
do not take any specific features of the patterns into account. For
patterns that can be considered as deformed periodic patterns a set of
disorder measures has been introduced that is sensitive to the
curvature and the deformation of the pattern \cite{GuHo98,HuNa05}. In
many cases the patterns are locally stripe-like. For such patterns
approaches have been developed that extract the local wavevector
\cite{EgMe98,CrMe94}, the orientation of the stripes \cite{HuEc95b},
and their curvature \cite{HuEc95b}. 

Many complex patterns exhibit striking, isolated objects like point
defects, e.g. dislocations \cite{ReRa89,DaBo02}, disclinations 
\cite{HaAd00} or penta-hepta defects \cite{CiCO90,YoRi03a}.
Dislocations often take on the form of spirals, as is the case for
points of vanishing oscillation amplitude in spatially extended
oscillations \cite{GiLe90,MaRi05b}. Similar spirals also arise in
excitable systems \cite{BaNe95,HiBa95}. Spiral-defect chaos in
convection simultaneously exhibits dislocations, disclinations, and
spirals \cite{MoBo93}. In systems displaying such point defects it has
proven useful to investigate the statistics of the defects or spirals
\cite{GiLe90,ReRa89,HiBa95,DaBo02,HuRi04} and also of their
trajectories \cite{GrRi01,HuRi04}.

Recently, a different type of characterization has been developed that
focuses on the topology of the patterns and extracts the number of
contiguous components and of their holes. It has been applied to 
numerical simulations of spatio-temporal chaos in excitable
reaction-diffusion models \cite{GaMi04} and to extensive experiments
on spiral defect chaos \cite{KrGaunpub}.

To identify specifically spirals and to measure their characteristics
only few methods have been used so far. In oscillatory systems in
which the spirals emanate from a defect with vanishing oscillation
amplitude the defect is relatively easily found and the number of
defects can be counted \cite{GiLe90}. In \cite{JeSp02} a fast method
is discussed to find the focus of spirals, independent of their
dynamical origin, which is based on the evolute of the emitted wave
pattern. In principle, spirals can also be detected from singularities
in the local orientation of the pattern \cite{HuEc95b} or the local
wavevector \cite{EgMe98}.

In this paper we discuss an approach that allows to extract typical
spiral features like the position of the spiral tip, the size of the
spiral measured in terms of the length of the spiral arm as well as its
winding number. The method is, however, not limited to proper spirals,
but characterizes arbitrarily shaped segments of stripe-like patterns
that have a typical wavelength. In the process other measures like the
number, size, and compactness of the components of the patterns are
obtained, as well. 

We discuss our diagnostic method at length in Sec.\ref{s:method}. In
Sec.\ref{s:analysis} we apply it to spiral defect chaos in thermal
convection and present the dependence of various measures on the
heating and on the Prandtl number of the fluid. We show that the
distribution function for the number of spirals in the pattern is
narrower than a Poisson distribution. The distribution of the winding
number of the spirals is reasonably approximated by an exponential as
had been suggested by experiments \cite{HuEc97}. We find that while
the decay rate depends only weakly on the heating it does depend
strongly on the Prandtl number. Correlating the contour length and the
area of the components we find the signature of target patterns for
$Pr=1.5$, but not for $Pr=0.3$. For $Pr=1.5$ the joint distribution
for the spiral length and the winding number reveals that Archimedean
spirals and relatively straight components dominate the patterns,
while for $Pr=0.3$ no such distinct groups are found. Instead, in this
regime the patterns are characterized by a large number of small
compact components. In Sec.\ref{s:conclusion} we discuss our
conclusions.

\section{Method}

\LB{s:method}


Our main goal is to identify spiral-like structures in a sequence of 
patterns and to characterize them quantitatively. In the process we
also obtain some other measures of the patterns. Here we describe the
procedure in detail. It is based on a set of contour lines giving the
pattern. Spiral arms are defined via the tip at the point of highest
curvature and the merging of the arm with other parts of the pattern
at a vertex (cf. Fig.\ref{f:spiral}b below).

We start from a sequence of snapshots that characterize some
continuous quantity of the system. In our example of convection it is
the temperature field in the mid-plane of the convection cell. While
in many pattern analyses such a continuous representation is first
reduced to a binary representation, we use the continuous
representation to obtain the set of contour lines corresponding to the
level midway between the minimal and the maximal value of the field
taken across all frames of the sequence. The resulting contour lines
retain more spatial resolution than the interface of a binary pattern.
Our simulations of the convection system employ periodic boundary
conditions. To reflect this periodicity, we appropriately connect
contours that reach the boundaries. Most contours become closed in
this process. In many patterns there are, however, also a few open
contours; they wrap around the whole system and may, for instance,
correspond to a straight convection roll that spans the system and
connects with itself across the periodic boundaries.

Having obtained all closed contours we already are in the position to
obtain a few simple quantitative characterizations of the pattern. The
number of closed contours provides a topological measure of the
pattern by giving its number of components. In many patterns it is
useful to distinguish between `white' and `black' components, e.g.
contours inside of which the temperature is higher (lower) than the
level of the contour. Focussing, for instance, on the pattern made up
of the white components, the number of white components gives then the
Betti number of order 1, while in our case of periodic boundary
conditions the number of black components is related to the Betti
number of order 2, which counts the number of holes in a pattern
\cite{GaMi04,KrGaunpub}. It is often useful to go beyond the
topological characterization of the patterns and also measure metric
aspects like the size of components. We therefore keep track also of
the length ${\mathcal P}$ of the perimeter of each component, i.e. the
length of the enclosing contour, and of its area ${\mathcal A}$.
Comparing the length and area of a given closed contour gives a
measure for the compactness of the component, which we define as  
\bea
{\mathcal C}=4\pi \frac{{\mathcal A}}{{\mathcal P}^2}. 
\eea 
With this normalization a circle has ${\mathcal C}=1$, while small
values of ${\mathcal C}$ correspond to filamentary structures.

To extract spirals from the contour lines we start by finding the tips
of the spirals, which we identify with the points of highest positive
(convex) curvature of the contour. In disordered patterns and, in
particular, in noisy patterns points with high curvature may also
appear away from the tips. To reduce the number of these spurious tips
we first smooth the contour. To do so we `diffusively' move the
polygon points $(x_i,y_i)$, $i=1\ldots N$, which define the contour,
using the mapping
\bea
x_i \rightarrow x_i+\delta (x_{i+1}-2x_i+x_{i-1})
\eea
and analogously for $y_i$. To avoid overshoots we use a small value of
$\delta $, typically $\delta=0.2$, and perform the smoothing
iteratively (typically in 50 steps). While the smoothing strongly
reduces localized bulges along overall straight sections of the
contour and therefore reduces the number of spurious tips,  the
curvature at a spiral tip is not affected much. Note that small
components shrink substantially by the smoothing. In this procedure we
are, however, mostly interested in large components and their
reduction in size is quite small. This is illustrated in
Fig.\ref{f:spiral}b which depicts a white component in the bottom left
quadrant of the pattern shown  of Fig.\ref{f:spiral}a before (green
open points) and after (red open circles) the smoothing. A further
benefit of the smoothing is that it distributes the grid points 
$(x_i,y_i)$ more equally along the contours. 

\begin{figure}
\epsfxsize=4cm {\epsfbox{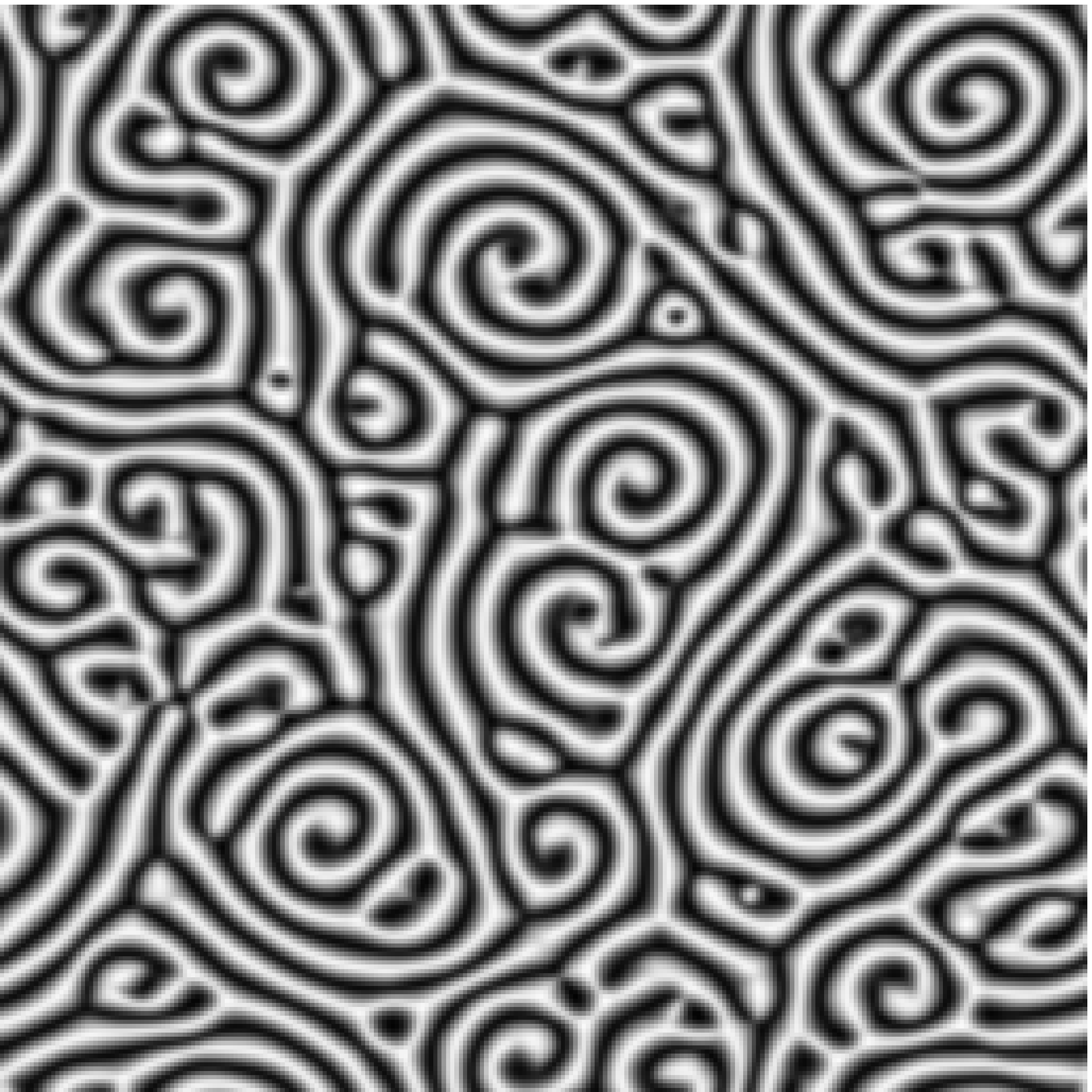}}
\hs{.2cm}
\epsfxsize=4cm {\epsfbox{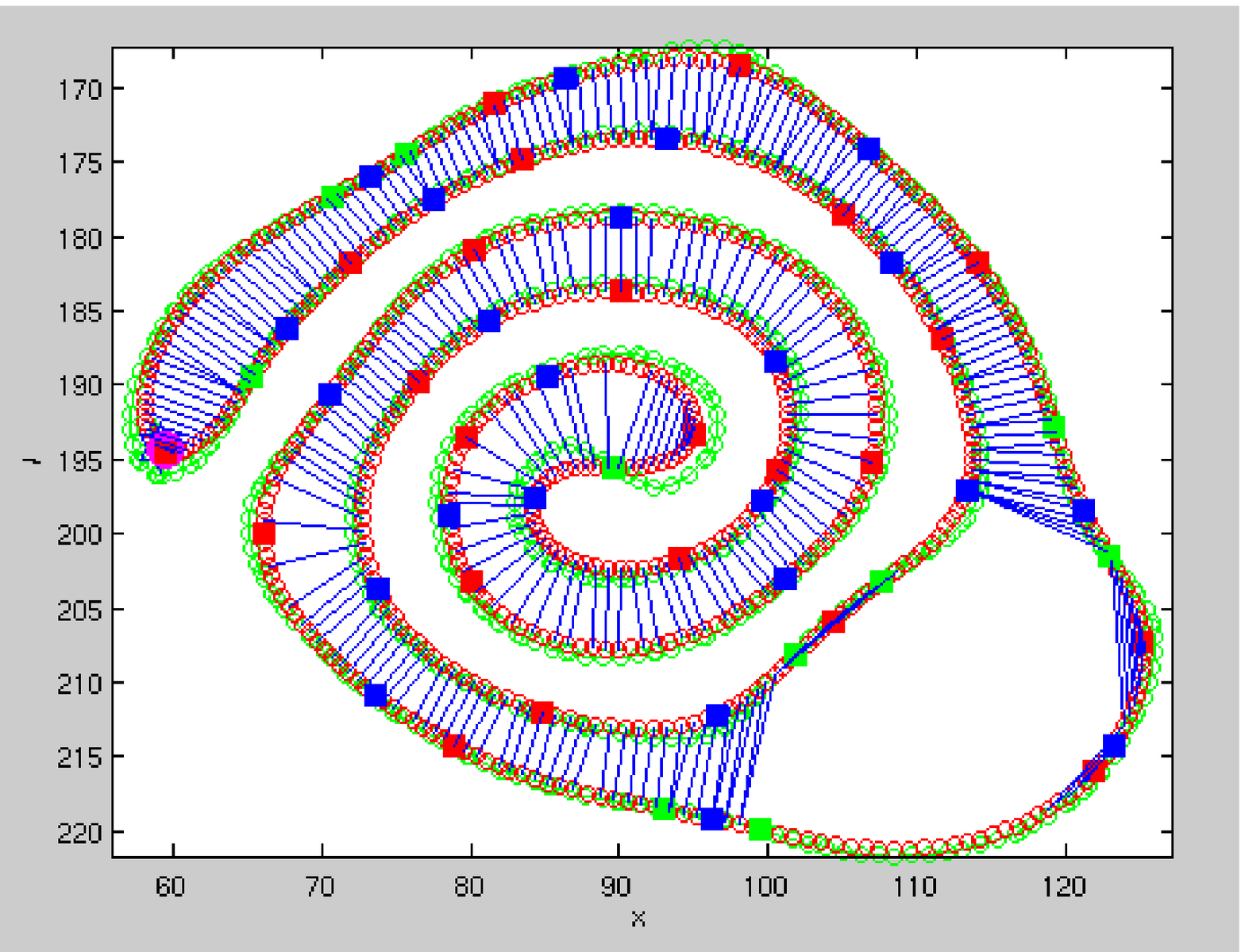}}
\caption{
a) Example of a convection pattern for $Pr=1.0$ and $\epsilon=1.0$ 
exhibiting spiral defect chaos.
b) Example of a component in the left-bottom quadrant of the pattern 
in a). Original contour (green open circles), smoothed contour (red
open circles), local maxima and minima of the curvature (red and blue
solid squares, respectively), inflection points (green solid squares),
shortest distance across the components (blue line). 
Here the axes are in terms of the grid coordinates rather than
physical coordinates.}
\LB{f:spiral}
\end{figure}

To obtain the local curvature of the contour we parametrize for each
point $(x_i,y_i)$ of the contour a small section of the contour
consisting of $5$ to $11$ points around $(x_i,y_i)$ by the arclength
$s$ and perform a best fit with two quadratic polynomials
$(x(s),y(s))$. The local curvature is then given by
\bea
\kappa_i =\frac{x'\,x''-y'\,y''}{\sqrt{x'^2+y'^2}^3},
\eea
where the primes denote first and second derivatives of $x(s)$ and 
$y(s)$, respectively, at the midpoint $(x(s_m),y(s_m))$ of the arc
approximating the contour over this local range. 

We then identify the spiral tips contained in this contour iteratively
by locating the point with maximal positive curvature. To measure the
length of this spiral arm we follow both branches of the contour
originating from the tip until the first vertex is reached (cf.
Fig.\ref{f:spiral}b). Vertices are identified by the width of the arm
exceeding a threshold $w_{max}$. Typically we take a value of
$w_{max}=10$, which corresponds to about one wavelength of the
pattern. To measure the width we find for each point on one branch
emanating from the spiral tip the closest point on the other. In
general this is a non-local problem. It becomes much simpler by
starting from the tip $(x_{i_{tip}},y_{i_{tip}})$ and finding for each
consecutive point $P_j^{(l)}\equiv (x_j^{(l)},y_j^{(l)})$,
$j<i_{tip}$, on the `lower' branch the point $P_k^{(u)}(P_j^{(l)})$ on
the `upper' branch that is closest to $P_j^{(l)}$  and that is within
a certain range $i_{tip}<k'\le k \le k'+\Delta k$ of the previously
identified point $P_{k'}^{(u)}(P_{j-1}^{(l)})$ on the upper branch. We
typically take $\Delta k=5$. In Fig.\ref{f:spiral}b the closest points
identified in this manner are connected by blue lines. Near the tip
this procedure sometimes leads to connections that lie outside rather
than inside the contour. To aid avoiding this problem, we require that
within a core region near the tip the point on the upper branch
strictly advance, $k>k'$. Typically we take the core region to be 10
points. When reaching a vertex, typically the points on the lower
branch still advance while those identified as closest on the upper
branch do not (e.g. near $x=115$ $y=197$ in Fig.\ref{f:spiral}b). We
then identify the end of the spiral arm with the last pair of points
$(P_j^{(l)},P_k^{(u)}(P_j^{(l)}))$ for which the point on the upper
branch still advanced, $P_k^{(u)}(P_j^{(l)}) \ne
P_{k'}^{(u)}(P_{j-1}^{(l)})$. For a segment or spiral arm thus
identified we take its length ${\mathcal S}$ to be the mean of the
arclength along its two branches and define its winding number via 
${\mathcal W}\equiv \frac{1}{2}(\theta^{(l)}+\theta^{(u)})/2\pi$,
where $\theta^{(l)}$ and  $\theta^{(u)}$ are the angles by which the two
branches of the contour are rotated from the tip to the vertex,
respectively.

Having found the first tip and the associated spiral arm we then
identify the second spiral tip with the point of maximal positive
curvature among the points on the contour not included in this first
arm and correspondingly for further arms. When identifying the
subsequent spiral arms we ensure that they do not overlap with
previously identified arms.  In many contours there are portions that
are not associated with a spiral tip. Nevertheless, even in these
parts of the contour the algorithm may identify segments that are
connected with local positive maxima of the curvature.
Fig.\ref{f:spiral}b shows such a segment in the lower right. It is
associated with a black inclusion. The size of the segments generated
in this way is of the order of $w_{max}$. To exclude such segments, we
omit all segments with an arclength smaller than $2w_{max}$ in the
final analysis of the results.

Many contours do not include a vertex. In those cases we associate the
contour with two segments (spiral arms) by cutting the contour at the
point $P_j^{(l)}$ (and $P_k^{(u)}(P_j^{(l)})$) at which the absolute
value of the curvature outside the core regions of the two tips is
minimal. 

The spirals of spiral defect chaos in convection are not topologically
stable objects; their winding number varies continuously and is not
conserved. In other systems spirals are topologically stable and can
only be created in pairs with opposite topological charge.
Well-studied systems of this type are oscillations in extended media
in which the spiral cores represent singularities of the phase of
oscillation at which the oscillation magnitude vanishes
\cite{GiLe90}.  In particular in such cases it is of interest to count
the number of spirals with positive and negative charge independent of
their size. Employing the algorithm discussed here, the topological
charge of the spirals can be related to the location of inflection
points of the contour in the vicinity of the spiral tip. Even in quite
small spirals the curvature changes sign near the tip and the side on
which this inflection point occurs depends on the topological charge
of the spiral. Note, however, that even segments that have a
noticeable winding number may not be classified as spirals according
to this criterion if their tip is part of a relatively straight
section. An example of such a segment is seen in Fig.\ref{f:spiral}b
where the segment with winding number ${\mathcal W}\sim 0.25$ is quite
straight near its tip and consequently the first inflection point
(marked with a green solid circle) is relatively far from the tip.

\section{Geometric Analysis of Spiral Defect Chaos}

\LB{s:analysis}

We use the approach discussed in Sec.\ref{s:method} to analyze
spiral defect chaos in Boussinesq convection \cite{MoBo93,BoPe00},
which is characterized by two system parameters: the Prandtl number
$Pr$ of the fluid, which measures the ratio between its viscosity and
its heat conductivity, and the heating, which is measured by the
dimensionless Rayleigh number $R$. The latter is often expressed in
terms of the scaled distance $\epsilon$ from the onset of convection
at $R_c=1708$, $\epsilon\equiv (R-R_c)/R_c$. 

Our results are based on direct numerical simulations of the
Navier-Stokes equations employing a pseudospectral code developed by
W. Pesch and co-workers \cite{DePe94a}. Details of the equations and
the code can be found in \cite{Pe96,YoRi03b}. The code is based on
Fourier modes in the horizontal direction and appropriate combinations
of trigonometric and Chandrasekhar functions that satisfy the top and
bottom boundary conditions in the vertical directions
\cite{Ch61,Bu67}. All runs are performed with $256\times 256$
horizontal Fourier modes for a system size of $L=32\cdot 2\pi/q_c$,
where $q_c=3.098$ is the critical wavenumber, and $2$ vertical modes.
Previous simulations have shown that this spatial resolution is
sufficient to capture the spiral defect chaos at the very least
semi-quantitatively \cite{DePe94a,Pe96}. Since the data analysis
requires relatively long runs, going to a significantly higher spatial
resolution is beyond our current computational means. To solve for the
time dependence the code uses a fully implicit scheme for the linear
terms, whereas the nonlinear parts are treated explicitly using a
second-order Adams-Bashforth method. The time step is typically taken
to be $t_v/500$, where $t_v$ is the vertical diffusion time.  In our
analysis we use snapshots of the temperature field in the mid-plane of
the convection cell that are typically taken every $2t_v$. We focus on
three different values of $\epsilon$ and $Pr$ each. The number of
snapshots evaluated range from 500 for $Pr=0.3$ to 2000 for $Pr=1.5$. 
Examples of patterns at $\epsilon=1.0$ are shown in
Fig.\ref{f:spiral}a for $Pr=1.0$ and in 
Fig.\ref{f:snap-pr0.3-pr1.5}a,b for $Pr=0.3$ and $Pr=1.5$.

\begin{figure}
\epsfxsize=4.cm {\epsfbox{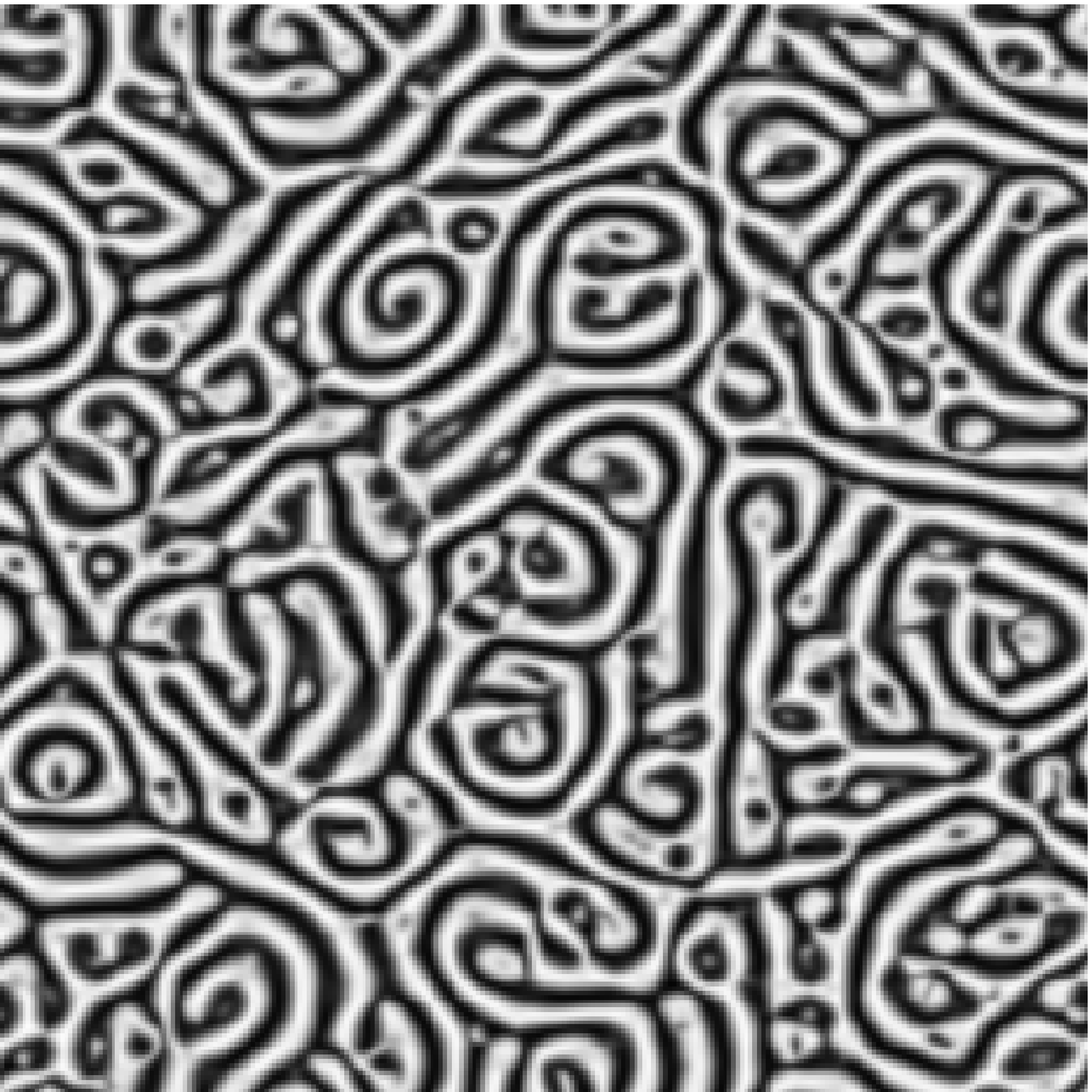}}
\hspace{.2cm}
\epsfxsize=4.cm {\epsfbox{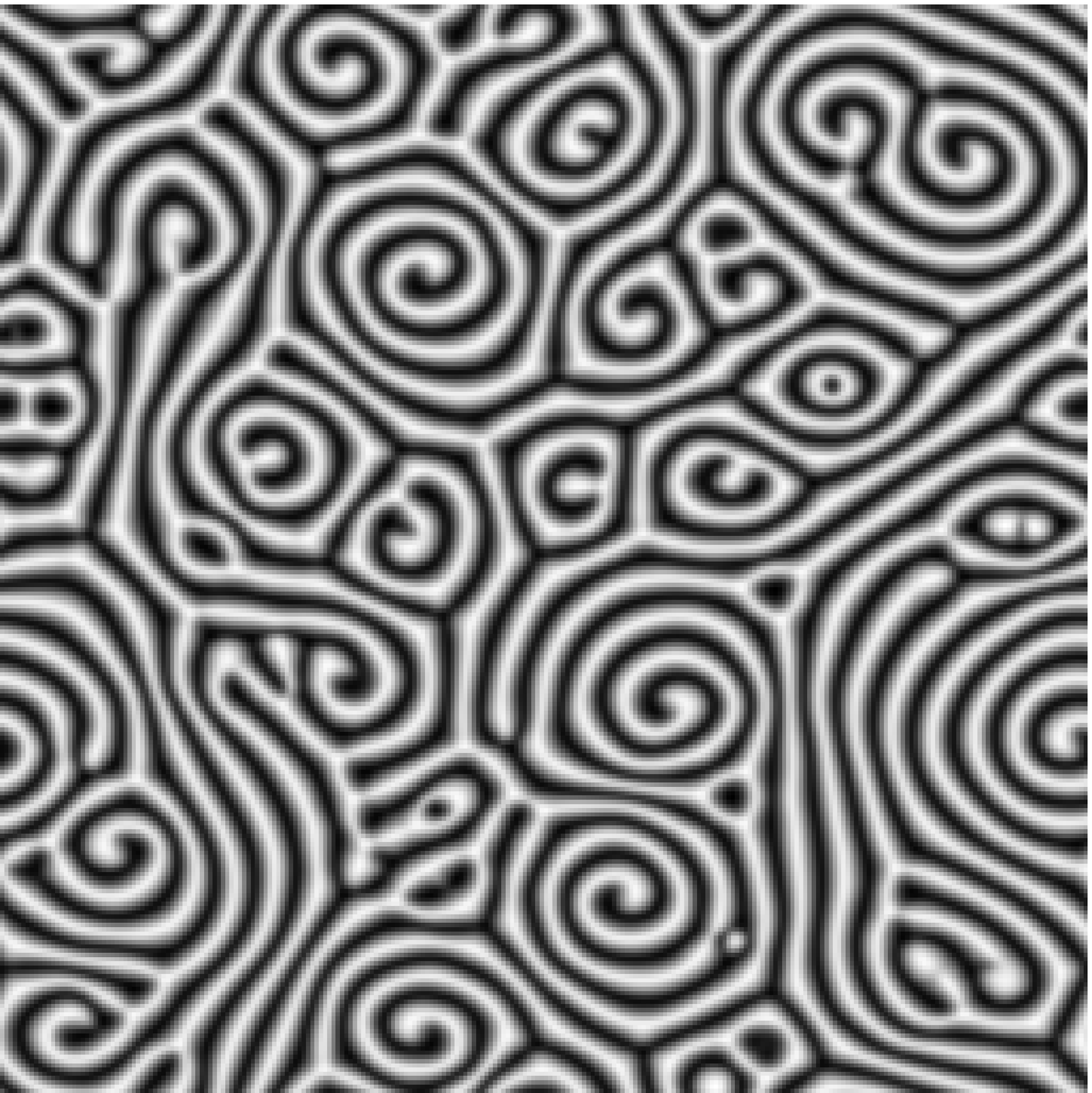}}
\caption{Snapshots of convection patterns at $\epsilon=1$ for a) $Pr=0.3$ and
b) $Pr=1.5$. System size $L=64.9$.}
\LB{f:snap-pr0.3-pr1.5}
\end{figure}

We first discuss the results for various mean quantities.
Fig.\ref{f:betti} shows the dependence of the mean number ${\mathcal
N}$ of closed contours on $\epsilon$ for three different Prandtl
numbers. More precisely, ${\mathcal N}$ gives the mean $({\mathcal
N}_b+{\mathcal N}_w)/2$ of `black' and `white' components. To assess
the variations in ${\mathcal N}$ over the course of the simulation
Fig.\ref{f:betti} shows in small symbols the mean values obtained from
averaging only over the initial, middle, and final third of the run.
In most cases the variations are so small that the additional symbols
are barely visible. The number of closed contours increases with
increasing $\epsilon$ and decreasing $Pr$. This is consistent with the
fact that the driving force of spiral defect chaos is a large-scale
flow, which is driven by the curvature of the convection rolls and
which increases with increasing $\epsilon$ and decreasing $Pr$. Such
an $\epsilon$-dependence has also been found in recent experiments for
$Pr\sim 1$ \cite{KrGaunpub}. There the Betti numbers of order 1 and 2
have been measured, which give the number of contiguous components and
of holes in the pattern, respectively \cite{GaMi04}. The mean length
of the contours exhibits the opposite behavior (data not shown). In
Fig.\ref{f:arclength} the analogous trend of increasingly finer
structures with increasing strength of the large-scale flow is
depicted in terms of the mean arclength of the segments (spirals) of
the components.

\begin{figure}
\epsfxsize=6cm {\epsfbox{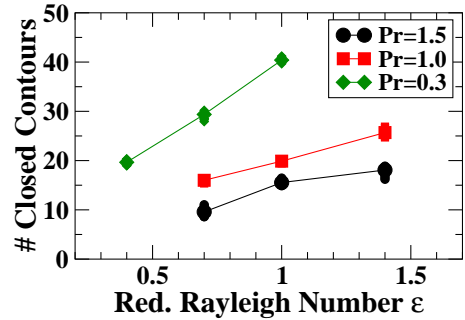}}
\caption{Mean number of closed contours as a function of $\epsilon$
for different values of the Prandtl number $Pr$.}
\LB{f:betti}
\end{figure}

The arclength results for $Pr=1.5$ show a large scatter at
$\epsilon=0.7$. This reflects the fact that for this system size
$\epsilon=0.7$ is close to the onset of spiral defect chaos
\cite{LiAh96}. Fig.\ref{f:snap-pr1.5eps0.7} shows that in  this regime
the pattern does not exhibit much of a spiral character. Our
computational resources do not allow to establish whether this low
chaotic activity persists indefinitely. Alternatively, the pattern
could eventually freeze in a disordered pattern. Experimentally, it
has been found that near the transition to spiral defect chaos the
system can exhibit very intermittent behavior in which long periods of
almost ordered patterns alternate with states exhibiting a few spirals
\cite{LiAh96}.

\begin{figure}
\epsfxsize=6cm {\epsfbox{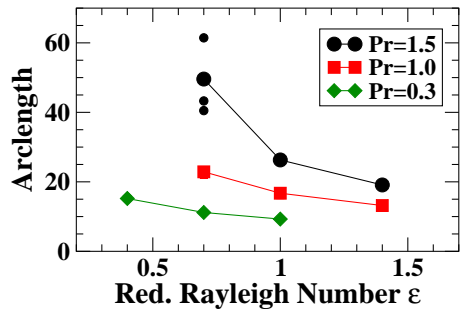}}
\caption{Mean arclength ${\mathcal S}$ of the segments of the
components as a function of $\epsilon$
for different values of the Prandtl number $Pr$.}
\LB{f:arclength}
\end{figure}

To extract the spiral character of the chaotic patterns we measure
their winding number. Since the system is chirally symmetric the mean
winding number is very close to 0 in all cases. The spiral character
is therefore better captured by the standard deviation of the winding
number. Its dependence on $\epsilon$ and $Pr$ is given in
Fig.\ref{f:sdwind}. Two trends can be discerned. For the strongly
chaotic states, i.e. for low Prandtl number and large $\epsilon$, the
standard deviation of the winding number decreases with increasing
$\epsilon$ and decreasing $Pr$, similar to the dependence of the
arclength. This reflects the fact that shorter spiral segments have a
smaller winding number. For low chaotic activity, however, the winding
number decreases with decreasing $\epsilon$, even though the 
arclength continues to increase with decreasing $\epsilon$. This is
very striking for $Pr=1.5$ and $\epsilon=0.7$ and consistent with the
visual appearance of the pattern shown in
Fig.\ref{f:snap-pr1.5eps0.7}. But even for $Pr=1.0$ the spiral
character of the state is non-monotonic in $\epsilon$. For these
parameter values no tendency for the pattern to become ordered was
apparent visually. Thus, the standard deviation of the winding number
may be suitable as a sensitive quantitative measure for the transition
to spiral defect chaos, complementing the previously used mean number
of spirals \cite{LiAh96}.

\begin{figure}
\epsfxsize=4cm {\epsfbox{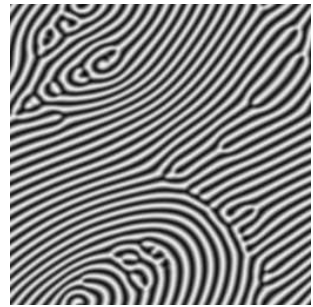}}
\caption{Snapshot of a typical convection pattern for $Pr=1.5$ and
$\epsilon=0.7$.}
\LB{f:snap-pr1.5eps0.7}
\end{figure}

A natural conjecture is that the dependence of the various measures on
$\epsilon$ and $Pr$ would be simpler if they were considered as
functions of the distance from the onset of spiral defect chaos,
$\epsilon-\epsilon_{SDC}(Pr)$. To test this in detail would require a
quantitative measurement of $\epsilon_{SDC}(Pr)$. This is beyond our
computational capabilities. However, while both the mean number of
components and the arclength take on comparable values for
$(Pr=0.3,\epsilon=0.4)$, $(Pr=1.0,\epsilon=1.0)$, and
$(Pr=1.5,\epsilon=1.4)$, this is not the case for the standard
deviation of the winding number, which reaches significantly higher
overall values for $Pr=1.5$ than for $Pr=0.3$. Thus, the values at
different Prandtl numbers cannot be related to each other by a simple
$Pr$-dependent shift and rescaling of $\epsilon$, indicating a
somewhat more complex scenario (see also
Figs.\ref{f:wind-dist-Pr1},\ref{f:wind-dist-eps1} below).

\begin{figure}
\epsfxsize=6cm {\epsfbox{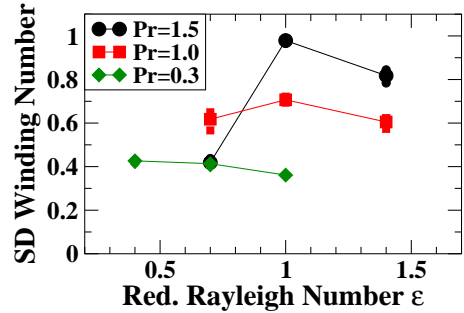}}
\caption{Standard deviation of the winding number as a function of $\epsilon$
for different values of the Prandtl number $Pr$.}
\LB{f:sdwind}
\end{figure}

In systems in which spirals constitute topologically conserved defects
it has proven quite instructive to measure the distribution function
for the number of spirals in a given snapshot
\cite{GiLe90,ReRa89,HiBa95,DaBo02,HuRi04}. In various, but not all
\cite{HiBa95,HuRi04} systems the distribution function was found to be
well fit by a squared Poisson distribution, which is consistent with a
simple over-all behavior of the spirals: they are created pairwise
with a fixed probability, diffuse randomly, and annihilate each other
pairwise upon collisions. The spirals in spiral defect chaos are not
topologically conserved and may exhibit different statistics. Indeed,
in a substantial effort the distribution function was measured
previously by visual inspection of many experimentally obtained
snapshots \cite{EcHu97}. There it was found for $Pr=0.96$ in a system
with aspect ratio $\Gamma=52$ that the probability distribution
function was consistent with a Poisson distribution for smaller values
of $\epsilon$ ($\epsilon=0.72$ and $\epsilon=0.80$), while for
$\epsilon=0.96$ the distribution was more sharply peaked than a
Poisson distribution. Of necessity, the manual counting limited the
number of snapshots that could be analyzed and the results have a
somewhat larger margin of uncertainty. In addition, the distribution
was determined only for spiral arms with a minimal winding number of
${\mathcal W}_{min}=1$. To address the question of the distribution
function more precisely and its dependence on the selection criterion
for the spirals we show in Fig.\ref{f:spiral-pdf} the distribution
function for three different values of the threshold, ${\mathcal
W}_{min}=\frac{1}{4}$,  ${\mathcal W}_{min}=\frac{1}{2}$, and
${\mathcal W}_{min}=1$ ($Pr=1.0$ and $\epsilon=1.0$). Our results
confirm with higher accuracy the experimental finding that the
distribution function is narrower than a Poisson distribution.
Moreover, shifting the distribution functions by their means and
rescaling them by the square root of their means, which would
correspond to the standard deviation for a Poisson distribution,
results in the distributions shown in Fig.\ref{f:spiral-pdf-rescaled}
(note the logarithmic scaling). Within the statistical error the
distribution functions collapse independent of the threshold
${\mathcal W}_{min}$ and are significantly more sharply peaked than
the Poisson distribution. So far no convincing model has been put
forward that provides an understanding of the form of these
distribution functions.

\begin{figure}
\epsfxsize=6cm {\epsfbox{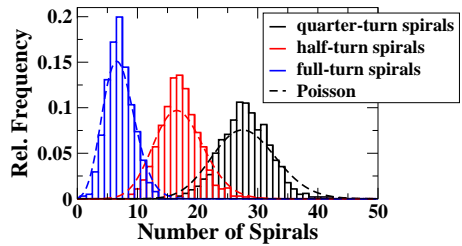}}
\caption{Distribution function for the number of spirals for three
different values of the threshold ${\mathcal W}_{min}$ for $Pr=1.0$
and $\epsilon=1.0$. Dashed lines
give fits to a Poisson distribution.}
\LB{f:spiral-pdf}
\end{figure}

\begin{figure}
\epsfxsize=6cm {\epsfbox{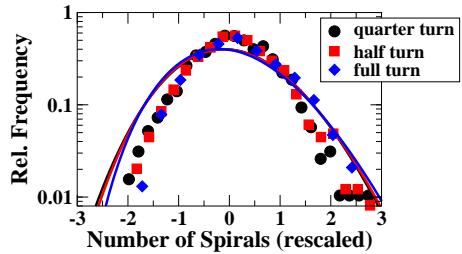}}
\caption{Rescaled distribution function for the number of spirals for the three
different values for the threshold ${\mathcal W}_{min}$. The curves are
obtained from those of Fig.\ref{f:spiral-pdf} by shifting and
rescaling by the mean of the distribution.}
\LB{f:spiral-pdf-rescaled}
\end{figure}

\begin{figure}
\epsfxsize=9cm {\epsfbox{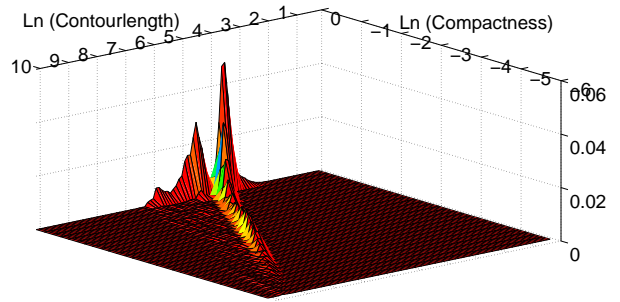}}
\epsfxsize=9cm {\epsfbox{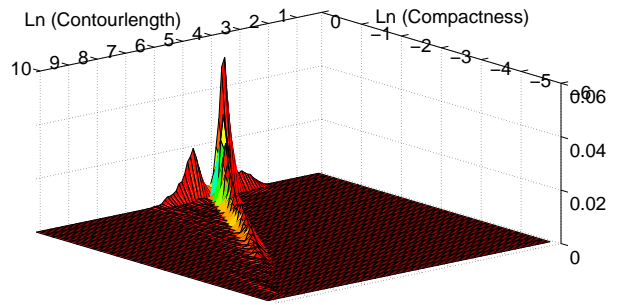}}
\epsfxsize=9cm {\epsfbox{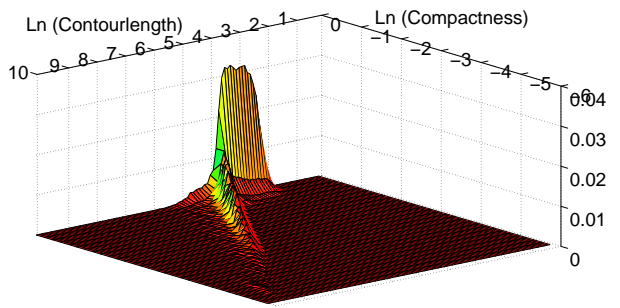}}
\caption{Joint distribution function for the logarithm of the compactness, 
$\ln {\mathcal C}$, and the logarithm of the contour length, 
$\ln {\mathcal P}$, for $\epsilon=1.0$. 
a) $Pr=1.5$, b) $Pr=1.0$, c) $Pr=0.3$.}
\LB{f:corr-cont-comp}
\end{figure}

Further insight into the patterns and their dependence on the system
parameters can be gained by considering correlations between various
properties. An instructive example is the joint distribution function
for the logarithm of the compactness of the closed contours and the
logarithm of their contour length, which is shown in
Fig.\ref{f:corr-cont-comp}. We find that most components have a
compactness ${\mathcal C}\sim 1$ and small contour length. For larger
contour lengths a relatively straight ridge is found in the
distribution function. It reflects the fact that the stripes have a
typical width $\lambda/2$. Thus, for large values of ${\mathcal P}$ the
compactness ${\mathcal C}$ can be approximated by
\bea
{\mathcal C}\sim 4\pi
\frac{\lambda \, {\mathcal P}}{4{\mathcal P}^2} \propto {\mathcal
P}^{-1}.
\eea
A few large components deviate from this relationship and have a
larger compactness; no component was found that has a smaller
compactness. This asymmetry arises because the wavelength of the
pattern cannot become smaller than a certain minimal value. For large
Prandtl number an additional structure is apparent: for values of the
compactness ${\mathcal C}\lesssim 1$  the distribution function 
shows three quite distinct peaks in its dependence on the
contour length. We associate these peaks with the appearance of target
patterns, which are comprised of a discrete set of quite compact
contour lines. When the Prandtl number is decreased these peaks become
less pronounced (cf. $Pr=1.0$ in Fig.\ref{f:corr-cont-comp}b) and
disappear for $Pr=0.3$ (Fig.\ref{f:corr-cont-comp}c). Instead the main
peak becomes quite broad and extends towards small contour lengths
indicating the appearance of relatively small compact components.
Comparing the distribution functions for different $Pr$ it is also
apparent that the relative frequency of large components decreases
noticeably with decreasing Prandtl number. 

Direct information about the spiral character of the patterns is
obtained by considering the correlation between the winding number and
the arclength of the segments. This is shown in
Fig.\ref{f:corr-arc-wind} for three Prandtl numbers at $\epsilon=1.0$.
Since the wavelength of the stripes is essentially fixed, the winding
number ${\mathcal W}$ of any spiral is expected to be limited by that
of an Archimedean spiral with the same arclength ${\mathcal S}$. A
simple computation gives 
\bea
{\mathcal S}&=&\pi {\mathcal W} \alpha \sqrt{4\pi^2{\mathcal W}^2+1}
-\frac{1}{2}\alpha \ln \alpha \nonumber \\
&&+\frac{1}{2}\alpha \ln \left(2 \pi \alpha {\mathcal W}
+\alpha \sqrt{4\pi^2{\mathcal W}^2+1}\right).
\LB{e:archi}
\eea
As Fig.\ref{f:corr-arc-wind} shows, eq.(\ref{e:archi}) fits the
extremal values of ${\mathcal W}$ quite well for all three Prandtl
numbers. The fitted value of $\alpha=1.8$ corresponds to a wavenumber
$q=2.2$ for the rolls making up the spiral, which is consistent with
the peak of the azimuthally averaged wavevector spectrum measured
experimentally \cite{BoPe00}. A cut through the distribution function
covering the range $10\le{\mathcal S}\le 12.5$ is shown in
Fig.\ref{f:corr-arc-wind-cut} for three values of $Pr$. For $Pr=1.5$
it has three peaks. The outer two correspond to near-Archimedean
spirals while the center one to quite straight structures.  As the
Prandtl number is decreased the triple-peak structure disappears and
no signature of the Archimedean spiral is left. As the arclength
${\mathcal S}$ is increased the distribution function drops off very
rapidly for $Pr=0.3$ and, as mentioned before, spirals with winding
numbers well above 1 are found only for large Prandtl numbers. 


\begin{figure}
\epsfxsize=7cm {\epsfbox{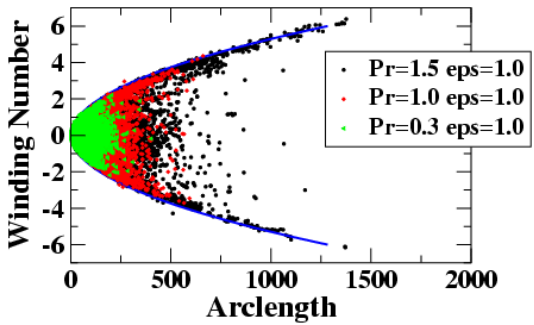}}
\caption{Correlation between arclength and winding number for
$\epsilon=1.0$. Each dot represents one segment of a component. The line is a fit to
an Archimedean spiral (cf. (\ref{e:archi})).}
\LB{f:corr-arc-wind}
\end{figure}

\begin{figure}
\epsfxsize=6cm {\epsfbox{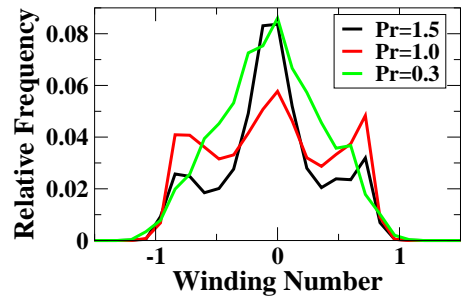}}
\caption{Distribution function of the winding number for arclengths
${\mathcal S}$ in the range $10 \le {\mathcal S} \le 12.5$
(cf. Fig.\ref{f:corr-arc-wind}).}
\LB{f:corr-arc-wind-cut}
\end{figure}

In experiments using CO$_2$ with $Pr=0.98$ it has been found that for
$\epsilon=0.72$ and $\epsilon=0.84$ the distribution function for the
winding number decays approximately exponentially
\cite{EcHu97}\footnote{In Fig.10 of \cite{EcHu97} the variable $m$
denotes phase windings of $\pi$ and not of $2\pi$.}. Due to the manual
counting of the spirals only a limited amount of data was available.
We have extracted this distribution function also from our data. In
Fig.\ref{f:wind-dist-Pr1} we show the distribution for the magnitude
of the winding number for three values of $\epsilon$ for $Pr=1.0$. In
particular for $\epsilon=0.7$ quite convincing exponential behavior is
found. With increasing $\epsilon$ the function decays, however,
somewhat faster at large winding numbers, casting some doubt on the
exponential behavior. Overall, there is, however, little change with
$\epsilon$ and fits to exponentials would lead to similar decay
constants. By contrast, when considering different values of the
Prandtl numbers quite different results are obtained as shown in
Fig.\ref{f:wind-dist-eps1}, where we keep $\epsilon$ fixed at
$\epsilon=1.0$. With increasing Prandtl number the decay of the
distribution function becomes significantly slower and spirals with
larger winding numbers become much more likely, as might have been
surmised from Fig.\ref{f:corr-arc-wind}. Thus, as became apparent
already in the discussion of
Figs.\ref{f:betti},\ref{f:arclength},\ref{f:sdwind}, the state of
spiral defect chaos depends on the Prandtl number and $\epsilon$
separately and not only on the distance $\epsilon-\epsilon_{SDC}(Pr)$
from its onset.

\begin{figure}
\epsfxsize=6cm {\epsfbox{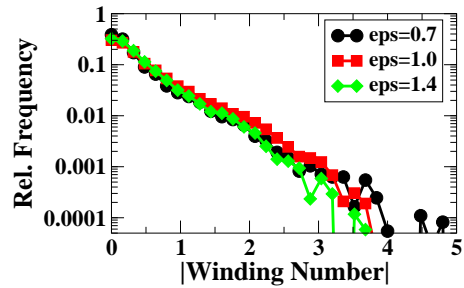}}
\caption{Distribution function of the winding number 
for $Pr=1.0$ and different values of $\epsilon$.}
\LB{f:wind-dist-Pr1}
\end{figure}

\begin{figure}
\epsfxsize=6cm {\epsfbox{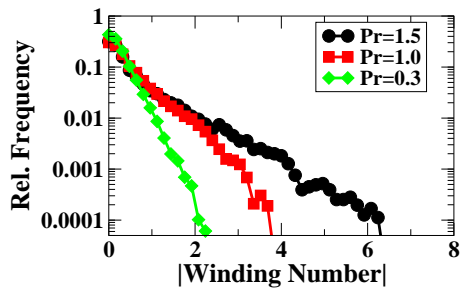}}
\caption{Distribution function of the winding number 
for $\epsilon=1.0$ and different values of $Pr$.}
\LB{f:wind-dist-eps1}
\end{figure}

Since the spirals break the chiral symmetry a particularly interesting
question is how they respond to an external breaking of the chiral
symmetry as it arises in a rotating convection system. This question
has been addressed previously in experiments using CO$_2$ at
$Pr=0.98$. There the number of clock-wise and counter-clockwise
spirals was counted by hand and it was found that the rotation
introduces an imbalance between the two populations \cite{EcHu95}. In
terms of the winding number we expect a smooth shift of the
distribution function with increasing rotation rate $\Omega$, as is
indeed borne out in Fig.\ref{f:dist-wind-rot}. The resulting
dependence of the mean winding number on the rotation rate is shown in
Fig.\ref{f:mean-wind-rot}.

\begin{figure}
\epsfxsize=6cm
{\epsfbox{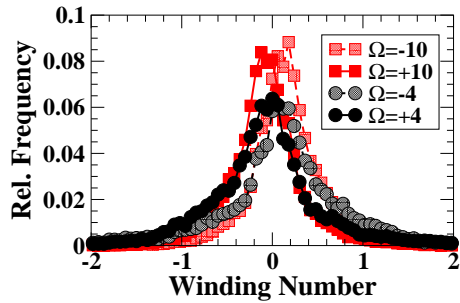}}
\caption{Breaking the chiral symmetry. Distribution function of the 
winding number for different rotation rates of the system
 ($Pr=1.0$ and $\epsilon=1.0$).}
\LB{f:dist-wind-rot}
\end{figure}

\begin{figure}
\epsfxsize=6cm
{\epsfbox{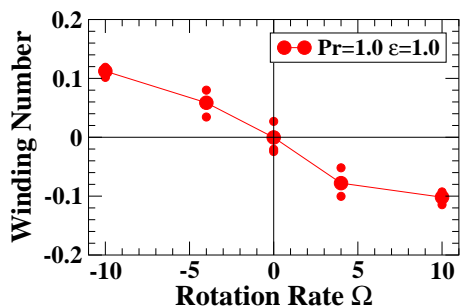}}
\caption{Dependence of the mean winding number on the rotation rate
for $Pr=1.0$ and $\epsilon=1.0$. Small symbols indicate the means over the
initial, middle, and final part of the run, respectively.} 
\LB{f:mean-wind-rot}
\end{figure}

As shown in Fig.\ref{f:mean-wind-rot}, with increasing $|\Omega|$ the
mean winding number decreases sublinearly. This is connected with the
fact that the segments of the pattern become shorter with increasing
rotation rate in anticipation of the K\"uppers-Lortz instability
\cite{KuLo69}. This decrease is shown in Fig.\ref{f:dist-arc-rot},
which presents the distribution function for the arclength for
different rotation rates. Note, that for $Pr\sim 1$ the onset of the
K\"uppers-Lortz instability is not reached until $\Omega \sim 14$
\cite{BoPe00}. If instead of the mean winding number the mean
topological charge of the patterns is plotted an essentially linear
behavior is found (data not shown).

\begin{figure}
\epsfxsize=6cm
{\epsfbox{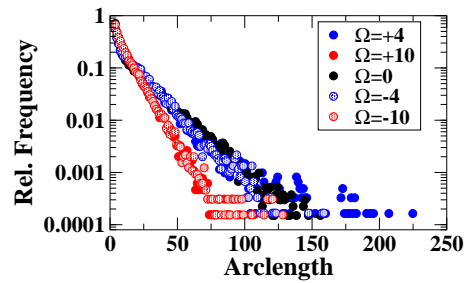}}
\caption{Distribution function of the arclength of the spiral segments
for different rotation rates ($Pr=1.0$ and $\epsilon=1.0$).}
\LB{f:dist-arc-rot}
\end{figure}

So far we have focussed in our analysis on the temperature field in
the midplane $z=0$ of the convective layer. While it seems reasonable
to assume that the pattern in the midplane is representative for the
convective patterns,  we test this assumption explicitly by measuring
the mean number of closed contours (`white' components, `black'
components, and the mean of the two), the mean number of spirals with
${\mathcal W}\ge 1$, the mean value of the arclength, and the standard
deviation of the winding number also at different heights $z$. For
these quantities Table \ref{t:z-dep} gives the ratio between the value
at the indicated $z$-value and the value at the midplane $z=0$, which
is given in the last column. In these runs we use a reduced system
size of $L=16 \cdot 2\pi/q_c$ retaining $128\times 128$ Fourier modes.
The Prandtl number is $Pr=0.8$.

\begin{table}
\begin{tabular}{|l|l|l|l|l|l|}
\hline 
&
z=-0.25&
z=0.25&
z=-0.125&
z=0.125&
z=0
\tabularnewline
\hline
\# Contours (`white')&
1.40&
0.733&
1.14&
0.876&
7.15
\tabularnewline
\hline
\# Contours (`black')&
0.738&
1.32&
0.896&
1.13&
7.28
\tabularnewline
\hline
\# Contours&
1.065&
1.031&
1.017&
1.002&
7.21
\tabularnewline
\hline
\# Spirals&
1.09&
1.11&
1.02&
1.06&
1.22
\tabularnewline
\hline 
Arclength&
0.996&
1.03&
0.993&
1.02&
48.07
\tabularnewline
\hline 
SD Winding Number&
1.01&
1.05&
1.001&
1.02&
0.512
\tabularnewline
\hline 
\end{tabular}
\caption{Dependence of the mean number of closed contours (positive,
negative, and average of the two (cf. Fig.\ref{f:betti})), mean number of spirals, the
arclength, and the standard deviation of the winding number on the
vertical position of the temperature field. The first four columns of the
table give the ratio between the quantity at the indicated $z$-level 
and at the midplane ($z=0$). The last column gives the value at 
the midplane ($Pr=0.80$, $\epsilon=1.4$).}
\LB{t:z-dep}
\end{table}

Considering the number of contours enclosing `white' components, one
notices a strong dependence on the height $z$ at which the temperature
field is measured. For instance, while at $z=-0.25$ the mean number of
white components is $40\%$ higher than at $z=0$, it is almost $30\%$
lower at $z=+0.25$. However, if one averages over the white and the
black components this variation is reduced to $6.5\%$ or less. The
$z$-dependence of the other quantities is similarly low. Since the
experimental visualization techniques measure some vertical average of
the temperature field one can expect that the difference between
experimentally determined quantities and the computationally obtained
ones remains below $10\%$. In the case of the arclength and the
standard deviation of the winding number the difference is expected to
be even smaller. For some quantities like the average number of closed
contours the variation with $Pr$ and $\epsilon$ over the range
considered is noticeably larger than these differences. On the other
hand, some quantities like the standard deviation of the winding
number vary with $\epsilon$ and $Pr$ only by $10\%$ or $20\%$. For
these quantities the direct comparison between the computation and
experimentally determined values may be not quite as conclusive.
However, even if different experimental or computational measurement
protocols should lead to results that differ quantitatively from each
other, we expect that even less marked trends like the
non-monotonicity of the standard deviation of the winding number (cf.
Fig.\ref{f:sdwind}) should be readily resolvable as long as the
protocol is kept the same when changing $\epsilon$ or $Pr$.

\begin{figure}
\epsfxsize=6cm {\epsfbox{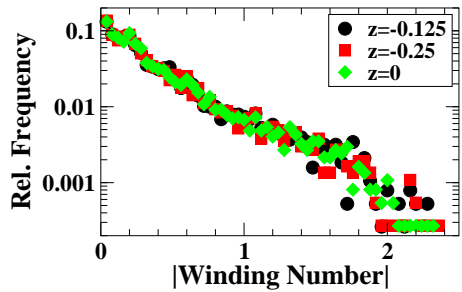}}
\caption{Dependence of the distribution function of the absolute value
of the winding number on the vertical position $z$ at which the 
temperature field is measured ($Pr=0.8$ and $\epsilon=1.4$).}
\LB{f:z-dep-win}
\end{figure}

As a further test of the significance of the $z$-dependence
Fig.\ref{f:z-dep-win} shows the distribution function for the winding
number at $z=0$, $z=-0.125$, and $z=-0.25$. Within the statistical
error these distributions are indistinguishable. We conclude therefore
that the measures discussed in this paper will be meaningful and will
provide valuable insight into these complex spiral patterns even
though the patterns themselves - and with them the measures - vary
somewhat across the layer.

\section{Conclusion}

\LB{s:conclusion}

Motivated by the fascinating convection patterns found in
spiral defect chaos we have introduced an approach to characterize
such patterns based on various features of the spirals. In the process
we have introduced other measures that are not specific to spiral-like
structures. We have used these various measures to characterize
spiral defect chaos as a function of the heating $\epsilon$ and the
Prandtl number $Pr$. 

Since the onset $\epsilon_{SDC}$ of spiral defect chaos depends on the
Prandtl number one could surmise that the properties of the resulting
spiral defect chaos state depend mostly on the distance
$\epsilon-\epsilon_{SDC}(Pr)$ from that threshold. Comparing the
$\epsilon$-dependence of the mean number ${\mathcal N}$ of components
of the pattern and of the arclength ${\mathcal S}$ of its segments
with the $\epsilon$-dependence of the standard deviation of the
winding number of the spirals we found evidence that this simple
relationship is, however, not satisfied. If any scaling relationship
should hold at all, it would have to involve also rescaling of the
dependent variables. Further support for this conclusion comes from
our finding that the distribution function of the winding number is
barely affected by changes in $\epsilon$, but it depends significantly
on the Prandtl number. 

We find that while the arclength of pattern segments increases
monotonically as the onset of spiral defect chaos is approached this
is not the case for the winding number of those segments. Thus,
further away from threshold the winding number increases with the
arclength as $\epsilon$ is decreased, indicating a lengthening of
spirals without much change in their structure. However, as the
threshold is approached the segments still become longer but their
winding number starts to decrease, i.e. the segments become
straighter. Possibly, the standard deviation of the winding number
could therefore serve as an early indicator for the transition to the
stationary state. 

With improved accuracy we confirm that the distribution function for
the number of spirals is narrower than a Poisson distribution
\cite{HuEc95}. Moreover, using different thresholds for the
classification of the spirals, we find that after rescaling these
different distribution functions collapse into a single curve. So far
no theoretical explanation for this distribution function is
available. 

Our automated analysis also allows higher accuracy for the measurement
of the distribution function of the winding number. In previous
experiments a roughly exponential behavior had been found
\cite{EcHu97}. Overall, we confirm this behavior, but at large winding
numbers the distribution functions typically decay somewhat more
rapidly. Interestingly, as mentioned above, the decay depends very
strongly on the Prandtl number, but very little on $\epsilon$. 

By correlating different measures we presented a more detailed
characterization of the patterns. For instance, correlating the
contour length of a component and its compactness, which is related to
its area, it is quite evident that for large Prandtl numbers the
patterns have many filamentary components and a sizeable number of
target patterns. For small Prandtl numbers the signature of the long
filaments and of the target patterns disappears and instead a much
broader spectrum of smaller, more compact components arises. In a
separate investigation \cite{MaRi05a} we find that the number of small
compact components strongly increases when the fluid properties depend
significantly on the temperature, i.e. when non-Boussinesq effects
become important. They introduce a resonant triad interaction between
the stripe modes that enhances the tendency towards hexagonal
(cellular) patterns. Of course, the breaking of the up-down symmetry
by the non-Boussinsesq effects leads also to differences between
closed contours that enclose up-flow rather than down-flow regions
(`black' and `white' components). 

Correlating the arclength with the winding number allows the
extraction of the spirals. For large Prandtl numbers not only the very
long components but also medium-sized components tend to be close to
Archimedean spirals. For small Prandtl numbers, however, even among
the longer components most have winding numbers that are much smaller
than the winding numbers of Archimedean spirals with the same
arclength. 

A strength of the measures employed in this paper is that they go
beyond purely local properties of the contour lines and identify
larger components making up the patterns. It should be noted, however,
that local changes that reconnect stripes can have large effects on
properties like the arclength and the winding number of the affected
pattern components. Moreover, in a given snapshot there are only few
large components. As a consequence of both aspects, these measures
fluctuate more strongly than purely local measures (e.g.
\cite{GuHo98}), necessitating a larger number of snapshots. In
particular in experiments this should pose no problem.

Since the approach discussed here identifies individual spirals one
would expect that it would be suitable to follow the evolution of
individual spirals dynamically. This would allow insight into the
mechanisms that set the mean size of the spirals and may help
understand the key elements that maintain the chaotic state. It has
been discussed that dislocation motion and wavenumber selection play
an important role \cite{CrTu95}. Visual inspection of the numerical
results indicates that spirals shrink largely through the compression
of their outer parts which then undergo a skewed-varicose instability
\cite{BoPe00}. 

We expect that the measures introduced here can be used beneficially
to investigate whether there are any transitions between different
states of spiral defect chaos as $\epsilon$ (or $Pr$) is changed. Such
an investigation would require, however, a much finer scan of the
parameter values, which is well beyond the reach of a computational
approach. Recently, such transitions have been reported based on
experimental investigations using the number of contours (Betti
numbers) as the sole measure \cite{KrGaunpub}. We expect that our
measures would provide detailed insight into the nature of these
transitions. At larger $\epsilon$ a transition from spiral defect
chaos to a chaotic state dominated by targets has been observed as the
Prandtl number is increased beyond $Pr=3.5$ \cite{AsSt93,AsSt94}. It
would be interesting to see whether the additional measures introduced
here shed further light on these aspects of spiral defect chaos. 

In conclusion, we have introduced a new set of measures for the
analysis of complex patterns, focusing on patterns that contain
spiral-like structures. Our results for spiral defect chaos suggest
that these measures should provide also insight into other complex
states that exhibit chiral structures. Examples include meandering
spirals and defect chaos in chemical systems
\cite{HiBa95,BeOu97,OuFl96,BaNe95}, spirals in vertically
vibrated material \cite{BrLe01}, patterns of bacterial colonies
\cite{LeBe04,Sh95}, calcium waves \cite{LeGi91a,HaZa98}, spiral waves
in the heart \cite{XuGu98} and in aggregating amoebae
\cite{PaLe97,DoKi01}.

We much appreciate the support by W. Pesch and his students, who
developed the code we have used for the computation of the patterns
\cite{DePe94a}. We also thank B. Winkler at the computing center of
the University of Bayreuth for maintaining the computer system used in
this work. We have benefitted from discussions with G. Ahlers, G.
Gunaratne, K. Krishan, K. Mischaikow, and M. Schatz. We gratefully
acknowledge support from the Department of Energy (DE-FG02-92ER14303)
and NSF (DMS-9804673).


\end{document}